\documentclass[12pt]{article}

\usepackage{amsfonts}

\usepackage{amsmath}
\usepackage{amssymb}
\usepackage{float}
\usepackage{titlesec}

\titleformat*{\section}{\large\bfseries}
\titleformat*{\subsection}{\bfseries}

\newtheorem{prop}{Proposition}[section]

\newtheorem{theorem}{Theorem}[section]

\newtheorem{lemma}[theorem]{Lemma}

 \def\2{I$\!$I}

\def\*{{\phantom *}}
\newcommand{\N}{\Bbb N}

\newcommand{\R}{\Bbb R}

\newcommand{\bml}[1]{\begin{multline}\label{#1}}
\newcommand{\bee}{\begin{equation}}
\newcommand{\bed}{\begin{displaymath}}
\newcommand{\ee}{\end{equation}}

\newcommand{\bs}{\begin{split}}

\newcommand{\ga}{\gamma} \newcommand{\Ga}{\Gamma}
 
\newcommand{\la}{\lambda}

\newcommand{\si}{\sigma}

\newcommand{\es}{\varnothing}
\newcommand{\w}{\widetilde}

\begin{document}
\large

\begin{center}
\renewcommand{\thefootnote}{\fnsymbol{footnote}}
{\Large \textbf{The new simplest proof of Ceyley's formula and connections with Kirkwood-Salzburg equations}}

\vspace{0.5cm}
{\textbf{ A.~L.~Rebenko}}
\date{}
\begin{footnotesize}
\begin{tabbing}
 Institute of Mathematics, Ukrainian National Academy of Sciences, Kyiv, Ukraine.
\end{tabbing}
\end{footnotesize}


\setcounter{footnote}{0} \renewcommand{\thefootnote}{\arabic{footnote}}


\vspace{0.5cm}
\textbf{Abstract} \\[0pt]
\end{center}
A new very simple proof of the number of labeled rooted forest-graphs with a given number of vertices is given.
As a partial case of this formula we have Cayley's formula.

\vspace{0.5cm}

\noindent \textbf{Keywords:} forests, trees, Cayley's formula.
\\
\noindent \textbf{Mathematics Subject Classification 2020:} 05C05, 82B31.

\medskip

\textheight 24.0cm \textwidth 16.0cm \headheight 0cm \headsep 0cm \footskip %
1cm \topmargin 0cm



\section{Introduction.}

\setcounter{equation}{0} \renewcommand{\theequation}{\arabic{section}.%
\arabic{equation}} 

There are many ways to prove the Cayley's formula(see, e.g. the books \cite{Moon67} and \cite{AiZi2010}).
About some recent publications see \cite{GuGu2017}.
A very close in style proof is also contained in the work \cite{TAK90}.
In this small note, we want to demonstrate another way to derive this formula. More exactly we establish a number
of rooted forest graph (i.e., graphs whose connected components are tree-graphs) with fixed number and positions of roots of their
connected components (tree-graphs). As a partial case of this formula we have Cayley's formula. It is important to note its connection
with cluster expansions in statistical mechanics. Although the formula for forests were obtained in another paper of the author, reference \cite{Re2021}, through the same way, here we have its derivation in the context of graph theory only and this shows an interesting contribution of how statistical mechanics ideas can provide new tools for another mathematics.

\section{Configuration space and forest graph.}

 Let ${{\Bbb R}}^{d}$ be a $d$-dimensional Euclidean
space. The set of coordinates $\gamma=\{x_i\}_{i\in{\Bbb N}} $ of identical
points is considered to be a locally finite subset in ${\Bbb
R}^d$ and the set of all such subsets is the configuration
space $\Gamma$. We will consider only finite graphs, so through $\Gamma_0$ denote the set of all finite configurations.
 Space of finite configurations in $ \R^d $ are possible
present in the form of disjunctive union of sets:
\begin{equation}\label{Ga0L}
\Ga_0\;:=\;\coprod_{n=0}^\infty \Ga^{(n)},\;\Ga^{(n)}\;:=\;\left\{ \ga\in\Ga\mid\; |\ga|\;=\;n,\;n\in\N \right\}, \;\;\Ga^{(0)}\;:=\;\es.
\end{equation}

One can form a graph from each configuration by connecting certain configuration points (vertices) with lines (edges).
A separate configuration point will also be considered as a graph consisting of a single vertex. If the graph $f$ is created
on the configuration $\gamma\in\Gamma^{(n)}$, then its order is determined by the cardinality of the configuration $|f|=|\gamma|=n $.
 By $E(f)$ we denote the set of edges of graph $f$.

Consider $ m $-connected graph $(m\geq 1)$, each connected component of which is labeled tree-graph.
The roots of the corresponding connected components form the configuration $ \eta = \{x_1, \ldots, x_m \} \in \Gamma^{(m)} $
All other vertices form some configuration $ \gamma = \{y_1, \ldots, y_n \}, \; n = 0, 1, ... $. Any subset of $\gamma$,
which belong to some tree with root $x_k$ could not belong to other tree with root $x_l, l\neq k$.
 The case $n=0$ means that $ m $-connected graph $f$ consist of $m$ points. In the case $m=1$ graph $f$ is labeled tree-graph with $n+1$ vertices
and $n$ edges. Such kind of graphs are called {\it rooted labeled graph-forest}. The set of all such forests with a given configuration of tree roots $ \eta $ and vertices
$ \gamma $ denote $\mathfrak{F}_{\eta;\gamma}$. Topologically, graphs corresponding to different configurations $\eta$ and $\gamma$
but with the same number of points $|\eta|$ and $|\gamma|$ are considered the same.

We consider analytic contributions of graphs as follows. Every vertex has analytic contribution some constant $h$ and for every edge
which connect point $x$ and $y$ we write some function $\nu(x-y)$. If such graphs appear in a specific problem, then it is necessary to impose appropriate conditions on these functions.

The analytical contribution of any graph-forest $f\in\mathfrak{F}_{\eta,\gamma}$ is
\begin{equation}\label{solQeqG}
 G(f)\;=\;h^{|\eta|+|\gamma|}\prod_{(x, y)\in E(f)}\nu(x-y),
\end{equation}

 The main result is based on the following elementary identity.
\begin{prop} For any $x\in\eta$ denote $\eta^{\hat{x}}:=\eta\setminus\{x\}$. Then
\begin{equation}\label{Edentity1}
\sum_{f\in\mathfrak{F}_{\eta;\gamma}}\prod_{(x', y')\in E(f)}\nu(x'-y')
=\sum_{\xi\subseteq\gamma}\prod_{(y\in\xi)}\nu(x-y)\sum_{f\in\mathfrak{F}_{\eta^{\hat{x}}\cup\xi;\gamma\setminus\xi}}\prod_{(x', y')\in E(f)}\nu(x'-y').
\end{equation}
\end{prop}
{\it Proof}. The left side of the equation \eqref{Edentity1} is the sum of the contributions of all graph-forests of the set $\mathfrak{F}_{\eta;\gamma}$.
The right side is the same sum in which its terms are written in the following order. First, we write the sum of the contributions of all graphs in which the vertex $x$ is not connected to any other vertex. On the right side this is the sum of all contributions of the graphs from $\mathfrak{F}_{\eta^{\hat{x}};\gamma}$,
that is $\xi=\emptyset$. The next group of terms includes graphs in which a vertex $x$ is joined by a single line to the vertices $y\in\gamma$. So, all $ \xi $ are
one point sets from $\gamma$. The remaining groups of sums corresponds to graphs in which the vertex $x$ is attached to the points of the subset $\xi$ of $\gamma$
by $|\xi|$ lines.

\hfill $\blacksquare$

\section{Reproducing kernel.}

\setcounter{equation}{0} \renewcommand{\theequation}{\arabic{section}.%
\arabic{equation}} 

For a given $\eta$ and $\gamma$ and factors $h, \nu$ we introduce some kernal $ Q_{h, \nu} (\eta | \gamma) $ which is uniquely determined by the following
recursion relation for any $x\in\eta$:
\begin{equation}\label{Qeq}
Q_{h,\nu}(\eta |\gamma)\;=\;h\sum_{\xi\subseteq\ga}
K_\nu(x;\xi)Q_{h,\nu}(\eta^{\hat{x}}\cup\xi |\gamma\setminus\xi),\,|\eta|\geq 1,
\end{equation}
where
\begin{align}\label{corf7Q}
K_\nu(x;\xi)\;:=\;\begin{cases}
 &\prod_{y\in\xi}\nu(x-y),\\
 & 1,\;\xi=\emptyset.
 \end{cases}
\end{align}
with the following initial conditions:
\begin{equation}\label{Qeq01}
Q_{h,\nu}(\emptyset |\emptyset)\;=\;1,\;\;Q_{h,\nu}(\emptyset |\gamma)\;=\;0,\;\;\text{if}\;\;\gamma\neq\emptyset,
\end{equation}
and
\begin{equation}\label{Qeq02}
Q_{h,\nu}(\eta |\gamma)\;=\;0, \;\text{if}\;\;\eta\cap\gamma\neq\emptyset.
\end{equation}

\begin{lemma}\label{proof1}
The solution of the equation \eqref{Qeq} can be written in the form:
\begin{equation}\label{solQeq}
Q_{h,\nu}(\eta |\gamma)\;=\;\sum_{f\in\mathfrak{F}_{\eta,\gamma}} G(f),
\end{equation}
where $G(f)$ is defined by \eqref{solQeqG}.
 \end{lemma}
The {\it proof} is trivial. It is easy to see that \eqref{solQeq} coincides with l.h.s. of \eqref{Edentity1}
and recursion relation \eqref{Qeq}-\eqref{corf7Q} is just r.h.s. of \eqref{Edentity1}.

 Now we need to know the number of
forest graphs for given  $ \eta, \gamma $ with $|\eta|=m$ and $|\gamma|=n$ This number is established by the following lemma (see also \cite{Re2021}).

\begin{lemma}\label{numberforest}\cite{Re2021}
The number of forest graphs in \eqref{solQeq}, which is a solution of \eqref{Qeq} with given
 $n=|\gamma|$ and $ m=|\eta|$ can be written by the following formula
\begin{equation}\label{numberG}
 N(m|n)\;=\;m(n+m)^{n-1}.
\end{equation}
 \end{lemma}

 {\it Proof. }If we put $ h = \nu = 1 $ in the equation \eqref{solQeq}, then $ Q_{1, 1}(\eta | \gamma) = N(m|n) $, which
satisfies the equation:
\begin{equation}\label{Numberforests}
N(m|n)=\sum_{k=0}^n {n \choose k}N(m+k-1|n-k).
\end{equation}
By induction in $n+m$ we will assume that the formula \eqref{numberG} is valid for
$ N(m + k-1|n-k) $ for $ k = 0, 1, \ldots, n $. Then substituting $ N(m + k-1|n-k) = (m + k-1) (m + n-1)^ {n-k-1} $ in the right part of \eqref{Numberforests}
we get the equality
\begin{equation}\label{Numberforests1}
\sum_{k=0}^n {n \choose k}(m+k-1)(m+n-1)^{n-k-1}= M_1\;+\;M_2,
\end{equation}
where
\begin{align}\label{M1}
M_1\;:=&\;m\sum_{k=0}^n {n \choose k}(m+n-1)^{n-k-1}=\\
=&m(m+n-1)^{-1}\sum_{k=0}^n {n \choose k}(m+n-1)^{n-k}=m(m+n-1)^{-1} (m+n)^{n},\nonumber \\
\end{align}
and
\begin{align}\label{M2}
M_2 :=& \sum_{k=0}^n {n \choose k}(k-1)(m+n-1)^{n-k-1}=\nonumber\\
 =& n\sum_{k=1}^n {n-1 \choose k-1} (m+n-1)^{n-k-1}-(m+n-1)^{-1} \sum_{k=0}^n {n \choose k} (m+n-1)^{n-k}=\nonumber \\
 =& n (m+n-1)^{-1}(m+n)^{n-1}-(m+n-1)^{-1}(m+n)^{n}\nonumber \\
 =&-m(m+n-1)^{-1}(m+n)^{n-1}.
\end{align}
As a result we have $M_1+M_2 \;=\;m(m+n)^{n-1} $ , which completes the proof.

\hfill $\blacksquare$

\section{Conclusion:\;Cayley's formula.}

So, for $m=1$ equation \eqref{numberG} is Cayley's formula. The main trick of this simple proof is to write the recursive equation \eqref{Qeq} for the kernel $ Q_{h, \nu}(\eta | \gamma) $.
This equation was written in the work \cite{MP77} as a technical element in writing the solution of the Kirkwood-Salzburg equations(see details in \cite{Re2021}).

\setcounter{secnumdepth}{0}

\vspace{0.5cm}

{\bf AVAILABILITY OF DATA}

Data available on request from the
author

\vspace{0.2cm}

\small{}

\end{document}